\documentclass[preprint,showpacs,preprintnumbers,amsmath,amssymb,eqsecnum]{revtex4}

\usepackage{graphicx}
\usepackage{dcolumn}
\usepackage{bm}
\usepackage{verbatim}

\def\z{{\mathbf{z}}}

\begin{document}


\title{Quantum Arrival and Dwell Times via Idealised Clocks}

\author{J.M.Yearsley}
\email{james.yearsley@imperial.ac.uk}

\author{D.A.Downs}
\email{delius.downs07@imperial.ac.uk}

\author{J.J.Halliwell}
\email{j.halliwell@imperial.ac.uk}

\author{A.K.Hashagen}
\email{anna-lena.hashagen07@imperial.ac.uk}

\affiliation{Blackett Laboratory \\ Imperial College \\
London SW7 2BZ \\ UK }

\begin{abstract}
A number of approaches to the problem of defining arrival and
dwell time probabilities in quantum theory make use of idealised
models of clocks. An interesting question is the extent to which
the probabilities obtained in this way are related to standard
semiclassical results. In this paper we explore this question
using a reasonably general clock model, solved using path integral
methods. We find that in the weak coupling regime where the energy
of the clock is much less than the energy of the particle it is
measuring, the probability for the clock pointer can be expressed
in terms of the probability current in the case of arrival times,
and the dwell time operator in the case of dwell times, the
expected semiclassical results. In the regime of strong
system-clock coupling, we find that the arrival time probability
is proportional to the kinetic energy density, consistent
with an earlier model involving a complex potential. We argue
that, properly normalized, this may be the generically expected
result in this regime. We show that these conclusions are largely
independent of the form of the clock Hamiltonian.
\end{abstract}



\maketitle

\newcommand\beq{\begin{equation}}
\newcommand\eeq{\end{equation}}
\newcommand\bea{\begin{eqnarray}}
\newcommand\eea{\end{eqnarray}}

\def\A{{\cal A}}
\def\D{\Delta}
\def\H{{\cal H}}
\def\E{{\cal E}}
\def\D{{\mathcal{D}}}
\def\p{\partial}
\def\la{\langle}
\def\ra{\rangle}
\def\ria{\rightarrow}
\def\Z{{\bf z}}
\def\t{{\tau}}
\def\y{{\bf y}}
\def\k{{\bf k}}
\def\q{{\bf q}}
\def\p{{\bf p}}
\def\P{{\bf P}}
\def\r{{\bf r}}
\def\d{{\partial}}
\def\s{{\sigma}}
\def\a{\alpha}
\def\b{\beta}
\def\e{\epsilon}
\def\z{\xi}
\def\U{\Upsilon}
\def\g{{\gamma}}
\def\l{{\lambda}}
\def\G{\Gamma}
\def\om{{\omega}}
\def\Tr{{\rm Tr}}
\def\iff{{\rm iff}}
\def\ih{{ \frac {i} { \hbar} }}
\def\trho{{\rho}}
\newcommand\bra[1]{\left<#1\right|}
\newcommand\ket[1]{\left|#1\right>}
\newcommand\brak[2]{\left<#1|#2\right>}
\newcommand\dif[2]{\frac{\partial #1}{\partial #2}}
\newcommand\diff[2]{\frac{\partial^{2} #1}{\partial #2^{2}}}

\def\au{{\underline \alpha}}
\def\bu{{\underline \beta}}
\def\pp{{\prime\prime}}
\def\id{{1 \!\! 1 }}
\def\half{\frac {1} {2}}

\def\h{\e} 

\def\jmy{james.yearsley@imperial.ac.uk}

\section{Introduction}

\subsection{Opening Remarks}

Questions involving time in quantum theory have a rich and
controversial history, and there is still much debate about their
status \cite{time, time2,MugLev,All}. Whilst historically most attention has
been focussed on tunneling times, because of their relevance to
atomic processes, there has more recently been considerable
interest in the problem of defining arrival and dwell times for
free particles. This shift in focus reflects the gradual
acceptance that the study of time observables in quantum theory is
as much a foundational issue as a technical one \cite{MugLev}.
Arrival and dwell times for a free particle are in some ways the
simplest time observables one could hope to define, and studying
these quantities allows one to see the difficulties common to all
time observables with the minimum of extra technical complication.
There are many different approaches to defining arrival and dwell
time probability distributions. In this paper we
use a clock model to define arrival and dwell times
and compare the results with standard semiclassical expressions.

\subsection{Arrival and Dwell Time}

We begin by reviewing some of the standard, mainly semiclassical,
formulae for arrival and dwell time. We consider a free particle
described by an initial wavepacket with entirely negative momenta
concentrated in $x>0$. The arrival time probability is the
probability $\Pi (t) dt $ that the particle crosses the origin in
a time interval $[t,t+dt]$.  A widely discussed candidate for the
distribution $\Pi (t)$  is the current density
\cite{MugLev,cur,HaYe1}:
\bea
\Pi(t)= J(t) &=& \frac{(-1)}{2m}\bra{\psi_{t}}
\left(  \hat{p} \delta(\hat x)+ \delta(\hat x)\hat{p}\ \right) \ket{\psi_{t}}\nonumber\\
&=&\frac{i}{2m}\left(\psi^{*}(0,t)\frac{\partial
\psi(0,t)}{\partial x}-\frac{\partial \psi^{*}(0,t)}{\partial
x}\psi(0,t)\right)\label{cur}
\eea
(We use units in which $\hbar = 1 $ throughout).
The distribution $\Pi(t)$ is
normalised to 1 when integrated over all time, but it is not
necessarily positive. (This is the backflow effect \cite{back}).
Nevertheless, Eq.(\ref{cur}) has the correct semiclassical
limit \cite{HaYe1}.

For arrival time probabilities defined by measurements, considered in this
paper, one might expect a very different result in the regime of
strong measurements, since most of the
incoming wavepacket will be reflected at $x=0$.  This is the
essentially the Zeno effect \cite{ Zeno}. It was found in a
complex potential model that the arrival time distribution
in this regime is the kinetic energy density
\beq
\Pi(t)=
C\bra{\psi_{t}}\hat p\delta(\hat x)\hat
p\ket{\psi_{t}}.\label{Zeno}
\eeq
where $C$ is a constant which
depends strongly on the underlying measurement model \cite{ech,Hal2}. (See Ref.\cite{KED} for a discussion of kinetic energy density.) 
However because
the majority of the incoming wavepacket is reflected, it is
natural to normalise this distribution by dividing by the
probability that the particle is ever detected, that is,
\bea
\Pi_N (t ) &=& \frac { \Pi ( t) } {  \int_0^\infty d s \Pi (s )  }
\nonumber \\
&=& \frac { 1 } { m | \langle p \rangle| }
\langle \psi_t |\hat p \delta ( \hat x ) \hat p
| \psi_t \rangle
\label{normalized}
\eea
where $ \langle p \rangle $ is the average momentum of the initial
state. This normalised
probability distribution does not depend on the details of the
detector. This suggests that the form
Eq.(\ref{normalized}) may be the generic result in this regime,
although a general argument for this is yet to be found.

The dwell time distribution is the probability $ \Pi (t) dt $ that
the particle spends a time between $[t,t+dt]$ in the interval
$[-L,L]$. One approach to defining this is to use the dwell time
operator,
\beq \hat
T_{D}=\int_{-\infty}^{\infty}dt  \chi(\hat x_{t})\label{dwellop},
\eeq
where  $\chi(x)$ is the characteristic function of the region
$[-L,L]$ \cite{time2a}. The distribution $\Pi (t)$ then is
\beq
\Pi(t)=\bra{\psi_{0}}\delta(t-\hat
T_{D})\ket{\psi_{0}}.\label{dwell} \eeq
In the limit $|p|L\gg1$, where $p$ is the momentum of
the incoming state, the dwell time operator has the approximate
form $ \hat T_{D}\approx {2mL } / {|\hat p|} $ so that the
expected
semiclassical form for the dwell time distribution is
\beq
\Pi(t)=\bra{\psi_{0}}\delta \left( t- \frac {2mL }  {|\hat p|}
\right)\ket{\psi_{0}}.
\label{dwellsemi}
\eeq

It is found in practice that measurement models for both
arrival and dwell time lead to
distributions
depending on both the initial state of the particle
and the details of the clock, typically of the form
\beq
\Pi_{C}(t)= \int_{-\infty}^{\infty} ds \ R (t,s) \Pi (s) \label{best},
\eeq
where $\Pi(t)$ is one of the ideal
distributions discussed above and the response function  $R (t,s) $
is some function of the clock
variables. (In some cases this expression will be a convolution).
However, it is of interest to coarse grain
by considering probabilities $p(t_1, t_2)$ for arrival or
dwell times lying in some interval $[t_{1},t_{2}]$. The resolution
function $R$ will have some resolution time scale associated with
it, and  if the interval $t_{2}-t_{1}$ is much larger than this
time scale, we expect the dependence on $R$ to drop out, so that
\beq
p(t_{1},t_{2})=\int_{t_{1}}^{t_{2}}dt\ \Pi_{C}(t)\approx \int_{t_{1}}^{t_{2}}dt\ \Pi(t)
\label{cg}
\eeq
This is the sense in which many different models are in agreement
with semi-classical formulae at coarse-grained scales.

Formulae such as Eqs.(\ref{cur}) and (\ref{dwellsemi}) and their
coarse grained version Eq.(\ref{cg})
are not fundamental quantum mechanical expressions,
but postulated semiclassical formulae. However, they have the
correct semiclassical limit and any approach to defining arrival
and dwell times must reduce to these forms in the appropriate
regime.

\subsection{Clock Model}

\begin{figure}[htbp] 
   \centering
   \includegraphics[width=5in]{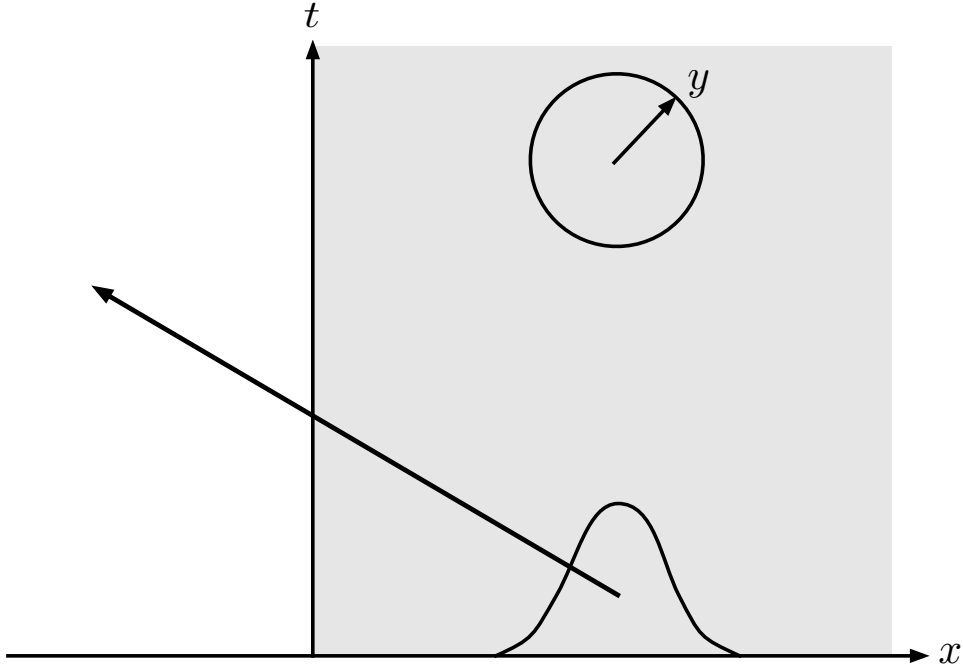}
   \caption{The arrival time problem defined using a model clock. The clock runs while the particle is in $x>0$.}
   \label{fig:1}
\end{figure}

In this paper we will derive arrival and dwell time
distributions by coupling the particle to a model clock. We denote the
particle variables by $(x,p)$ and those of the clock by
$(y,p_{y})$.  We denote the initial states of the particle and clock
by $\ket{\psi}$, $\ket{\phi}$, respectively, and the total system state by $\ket{\Psi}$.
We couple this clock to the particle via the
interaction $H_{I}=\l  \chi(\hat x) H_C$. The total Hamiltonian
of the system plus clock is therefore given by,
\beq
H=H_{0}+\l  \chi(\hat x) H_C \label{clk}
\eeq
where $H_{0}$ is the Hamiltonian of the particle. Here $\chi$ is
the characteristic function of the region where we want our clock
to run, so that $ \chi (x) = \theta (x)$ for the arrival time
problem and $\chi (x) = \theta (x+L ) \theta (L-x)$ for the
dwell time problem.
The operator,
\beq
H_{c}=H_{c}(\hat y, \hat p_{y})
\eeq
describes the details of the dynamics of the clock and we assume
that it is such that the clock position $y$ is the measured time.
The physical situation is depicted in Figure (\ref{fig:1}).
For the moment we will assume only that the clock Hamiltonian is
self adjoint, so that it may be written in the following form,
\beq
H_{c}=\int \ d\h \ \h\ket{\h}\bra{\h}
\eeq
where the $\ket{\h}$ form an orthonormal basis for the Hilbert
space of the clock. Later on we will restrict $H_{c}$ further by
considering the accuracy of the clock. We will also quote some
results for the special choice of $H_{c}=\hat p_{y}$, whose action
is to simply shift the pointer position $y$ of the clock in
proportion to time. This is the
simplest and most frequently used choice for the clock
Hamiltonian.
The physical relevance of this and other clock models is discussed
in Refs. \cite{clockchapter, clock}.

Our aim, for both arrival and dwell times,
is to first solve for the evolution of the combined system of
particle and clock. We write this as
\bea
\Psi(x,y,\t)&=&\bra{x,y}e^{-iH\t}\ket{\Psi_{0}}
\nonumber \\
&=&\bra{x,y}\exp\left(-iH_{0}\t-i\l \chi(\hat x)H_{c}\t\right)\ket{\psi_{0}}\ket{\phi_{0}}\nonumber\\
&=&\int d\h\brak{y}{\h}\brak{\h}{\phi_{0}}\bra{x}\exp(-iH_{0}\t-i\l \chi(\hat x)\h \t)\ket{\psi_{0}}\label{psifinal}
\eea
and we then solve for the propagator in the integrand using path
integral methods. We will take the total time $\t$ to be
sufficiently large that the wave packet has left the region
defined by $\chi (x)$.
We will then compute the final distribution of the pointer
variable $y$, which is
\beq
\Pi(y)=\int_{-\infty}^{\infty} dx|\Psi(x,y,\t)|^{2}\label{pred}.
\eeq
Our main aim is to show that the predictions of the clock model
Eq.(\ref{pred}) reduce, in certain limits, to the standard forms
described above.

\subsection{Connections to Earlier Work}

Clock models of the type Eq.(\ref{clk}) for arrival and dwell
times have been studied by  numerous previous authors, including
Peres \cite{clock}, Aharanov et al
\cite{clock2}, Hartle \cite{Hartleclock} and Mayato et al
\cite{clockchapter}. These studies are largely focused on the
characteristics of clocks.  Refs.\cite{clock,clock2} are the works
perhaps most closely related to the present work. They  concentrate 
on the case of a clock Hamiltonian linear in momentum, with some elaborations
on this basic model in the case of Ref.\cite{clock2}. Here, we focus on a different issue
not addressed by these works, namely
the dependence of the distribution Eq.(\ref{pred}) on
the initial state of the particle, for reasonably general clock
Hamiltonians. In particular, we determine the extent to which the standard
semiclassical forms derived above are obtained for general initial
states of the particle.
We also use path integral methods to perform the calculations,
in contrast to the scattering methods used in most of the previous
works. Path integral methods similar to those employed here have previously been used in Refs. \cite{Sok, Fer} to explore the time taken to tunnel under a potential barrier, although these authors sought to define the tunneling time in terms of subsets of paths in the path integral, rather than by considering the behavior of a physical clock.

\subsection{This Paper}

The rest of this paper is arranged as follows. In Section
\ref{sec2} we review some path integral techniques and in
particular the path decomposition expansion (PDX), which we
will use to compute Eq.(\ref{psifinal}). We also introduce a
useful semi-classical approximation. In Section \ref{sec3}
we compute Eqs.(\ref{psifinal}), (\ref{pred}) for the arrival
time problem, and similarly in Section \ref{sec4} for the dwell time problem.
We conclude in Section \ref{con}.

\section{The path decomposition expansion and the semiclassical approximation}\label{sec2}

In this section we discuss the path decomposition expansion (PDX),
which we shall make use of in the rest of this paper to calculate
Eqs.(\ref{psifinal}),(\ref{pred}). We also introduce a useful
semiclassical approximation, which significantly reduces the
complexity of the calculations. Throughout this section we shall
focus on the arrival time case, so that $\chi(x)=\theta(x)$.

To evaluate Eq.(\ref{psifinal}) we need to evaluate a propagator of the form
\beq
g(x_1, \tau | x_0 ,0 ) = \langle x_1 | \exp  \left( -  i (H_0 +V \theta(\hat x)) \tau \right)
| x_0 \rangle,
\label{12}
\eeq
for  $x_1<0$ and $ x_0 > 0$ (more general situations are considered in \cite{HaYe1}). Here $V$ is some real number. This may be calculated using a sum over paths,
\beq
g(x_1, \tau | x_0,0 ) = \int {\cal D} x \exp \left( iS\right)\label{22}
\eeq
where
\beq
S=\int_0^{\tau} dt \left( \half m \dot x^2 - V \theta (x) \right),
\eeq
and the sum is over all paths $ x(t)$ from $x(0) = x_0$ to $x(\tau) = x_1$.

\begin{figure}[htbp] 
   \centering
   \includegraphics[width=5in]{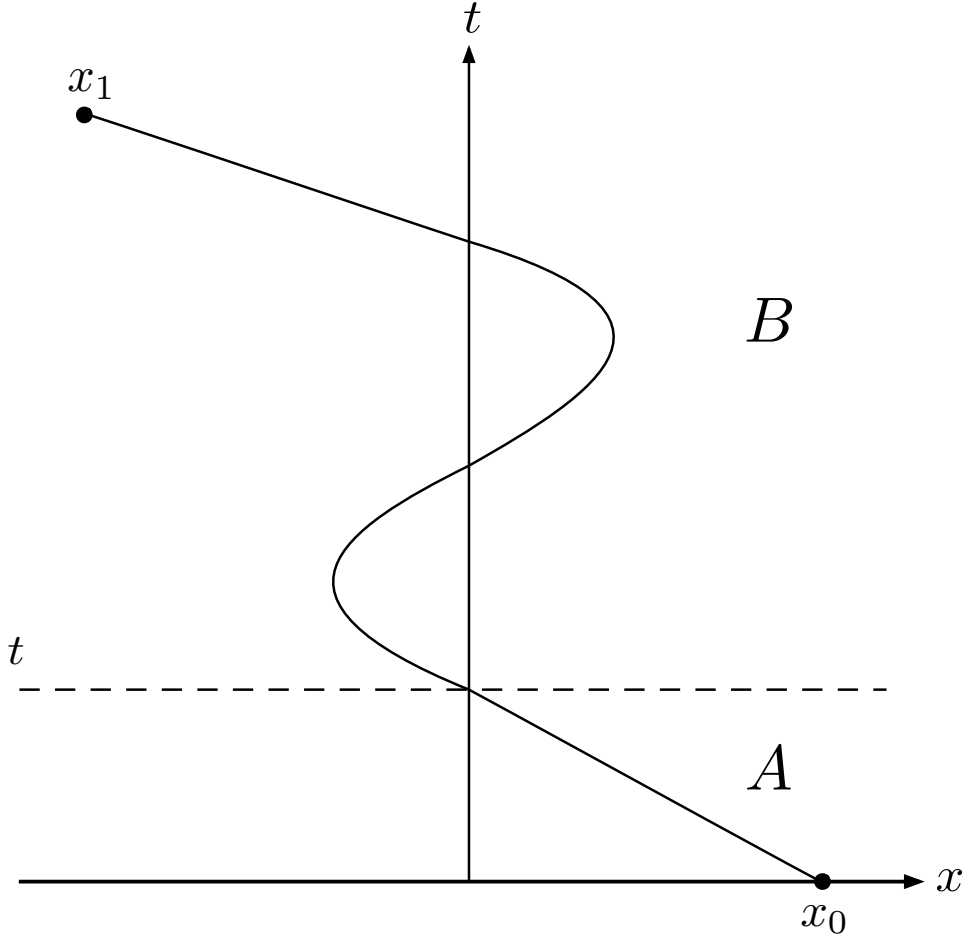}
   \caption{The first crossing path decompositon expansion: paths from $x_{0}>0$ to $x_{1}<0$ consist of a restricted section of propagation to $x=0$, $(A)$, followed by unrestricted propagation along $x=0$ and to $x_{1}<0$ $(B)$.   }
   \label{fig:2}
\end{figure}

We can simplify the analysis by splitting off the sections
of the paths lying entirely in $x>0$ or $x<0$. The way to do this is to use
the path decomposition expansion \cite{PDX,HaOr}.
Every path from $x_0>0$ to $x_1 <0$ will typically cross $x=0$ many times, but all paths have a first crossing, at time $t$, say. As a consequence
of this, it is possible to derive the formula,
\beq
g(x_1, \tau | x_0,0 ) =
\frac {i } {2m} \int_{0}^{\tau} dt
g (x_1, \tau | 0, t) \frac {\partial g_r } { \partial x} (x,t| x_0,0) \big|_{x=0}
\label{PDX1.1}
\eeq
Here, $g_r (x,t|x_0,0)$ is the restricted propagator obtained by a path integral of the form Eq.(\ref{22})
summed over paths lying entirely in $x>0$. Its derivative at $x=0$  is given by a sum over all paths in $x>0$ which end on $x=0$ \cite{HaOr}.
Similar formulas may also be derived involving the last crossing time, and both the first {\it and} last crossings times \cite{PDX, HaOr}.
The PDX is depicted in Figure (\ref{fig:2}).

Each element of these expressions can be calculated for a potential of the form $ V \theta (x)$.
The restricted propagator in $x>0$ is  given by the method of images expression,
\beq
g_r (y, \t |x,0) = \theta (y) \theta (x)
\left( g_f (y, \t |x,0) - g_f (-y, \t |x,0) \right)e^{iV\t},
\label{17}
\eeq
where $g_f $ denotes the free particle propagator
\beq
g_f (y, \t |x,0) = \left( \frac {m} {2 \pi i \t } \right)^{1/2}
 \exp \left( \frac {i m (y - x)^2 } { 2  \t} \right).
\eeq
Note that this means that
\beq
\left.\frac{\partial g_{r}}{\partial x}\right|_{x=0}=2\left.\frac{\partial g_{f}}{\partial x}\right|_{x=0},
\eeq
and thus Eq.(\ref{PDX1.1}) can be written as,
\beq
\bra{x_{1}}e^{-i(H_{0}+V\theta(\hat x))\t}\ket{x_{0}}=\frac{1}{m}\int_{0}^{\t}dt\bra{x_{1}}e^{-i(H_{0}+V\theta(\hat x))(\t-t)}\delta(\hat x)\hat p e^{-i(H_{0}+V)t}\ket{x_{0}}\label{PDX1},
\eeq
where $\delta(\hat x)=\ket{0}\bra{0}$ and $ \ket{0}$ denotes a
position eigenstate $ \ket{x}$ at $ x=0$.

The propagator from $x=0$ to $x_{1}>0$ is more difficult to calculate, because it will generally involve many re-crossings of the origin. This propagator may be calculated exactly by using the last crossing version of the PDX \cite{HaYe1}, but it may also be approximated using a semiclassical expression, which we now describe.

The exact propagator from the origin to a point $x_{1}<0$ consists of  propagation along the edge of the potential followed by restricted propagation from $x=0$ to $x_{1}$.
However, for sufficiently small $V$, we expect from the path integral representation of the propagator that the dominant contribution will come from paths in the neighbourhood of the straight line path from  $x=0$ to $x_{1}<0$. These paths lie almost entirely in $x<0$, so we expect that the propagator may be approximated semiclassically by
\beq
\bra{x_{1}}e^{-i(H_{0}+V\theta(\hat x))t}\ket{0}\approx \bra{x_{1}}e^{-iH_{0}t}\ket{0},
\eeq
and thus Eq.(\ref{PDX1}) can be written as,
\beq
\bra{x_{1}}e^{-i(H_{0}+V\theta(\hat x))\t}\ket{x_{0}}\approx\frac{1}{m}\int_{0}^{\t}dt\bra{x_{1}}e^{-iH_{0}(\t-t)}\delta(\hat x)\hat p e^{-i(H_{0}+V)t}\ket{x_{0}}\label{PDX2}.
\eeq
In Ref.\cite{HaYe1} it was shown that this semiclassical approximation holds for $E\gg V$, where $E$ is the kinetic energy of the particle.

\section{Arrival time distribution from an idealised clock}\label{sec3}

We now turn to the calculation of the arrival time distribution, Eq.(\ref{pred}),
recorded by our model clock. Using the path decomposition expansion in the form Eq.(\ref{PDX1})
the state of the system Eq.(\ref{psifinal}) can be written as,
\bea
\Psi(x,y,\t)&=&\bra{x,y}e^{-i(H_{0}+\l \theta(\hat x) H_{c})\t}\ket{\Psi_{0}}\nonumber\\
&=&\frac{1}{m}\int d\h\brak{y}{\h}\brak{\h}{\phi_{0}}\nonumber\\
&&\times\int_{0}^{\t}dt \bra{x}\exp(-i(H_{0}+\l \h\theta(\hat x))(\t-t))\delta(\hat x)\hat p\exp(-i(H_{0}+\l \h)t)\ket{\psi_{0}}.\label{atstart}
\eea
We can  simplify this expression in two different regimes,
the weak coupling regime of $E\gg\l \h$, and the strong coupling regime of $E\ll\l \h$.

\subsection{Weak Coupling Regime}\label{sec3.1}

In the limit $E\gg\l \h$ we can make use of the semiclassical approximation to the PDX formula, Eq.(\ref{PDX2}). This yields,
\bea
\Psi(x,y,\t)&=&\frac{1}{m}\int d\h\brak{y}{\h}\brak{\h}{\phi_{0}}\nonumber\\
&&\times\int_{0}^{\t}dt \bra{x}\exp(-iH_{0}(\t-t))\delta(\hat x)\hat p\exp(-i(H_{0}+\l \h)t)\ket{\psi_{0}}.\label{begin}
\eea
This means that the arrival time distribution is
\bea
\Pi(y)&=&\frac{1}{m^{2}}\int d\h d\h'\brak{\phi_{0}}{\h'}\brak{\h'}{y}\brak{y}{\h}\brak{\h}{\phi_{0}}\nonumber\\
&&\times \int_{0}^{\t}dt dt'\bra{\psi_{0}}\exp(i(H_{0}+\l \h')t')\hat p \delta(\hat x)\exp(-iH_{0}(t'-t))\delta(\hat x)\nonumber\\
&&\times\hat p \exp(-i(H_{0}+\l \h)t)\ket{\psi_{0}}\label{3.3}
\eea
(Recall that we are assuming $\t$ is sufficiently large that all the wavepacket is in $x<0$ at the final time, see the discussion below Eq.(\ref{psifinal}).) To proceed, we first note that for any operator $\hat A$, we have
 \beq
 \delta(\hat x)\hat A\delta(\hat x)=\delta(\hat x)\bra{0}\hat A\ket{0}.
 \eeq
Using this in Eq.(\ref{3.3}) gives,
\bea
\Pi(y)&=&\frac{1}{m^{2}}\int d\h d\h'\brak{\phi_{0}}{\h'}\brak{\h'}{y}\brak{y}{\h}\brak{\h}{\phi_{0}}\nonumber\\
&&\times \int_{0}^{\t}dt dt'\bra{\psi_{0}}\exp(i(H_{0}+\l \h')t')\hat p \delta(\hat x)\hat p \exp(-i(H_{0}+\l \h)t)\ket{\psi_{0}}\nonumber\\
&&\times\bra{0}\exp(-iH_{0}(t'-t))\ket{0}\label{3.5}.
\eea
We see here the appearance of the combination $\hat p \delta(\hat x)\hat p$, and the main challenge is to show how this turns into the current operator, $\delta(\hat x)\hat p + \hat p \delta(\hat x)$.

Next we rewrite the integrals using
\beq
\int_{0}^{\t}dt dt'=\int_{0}^{\t}dt\int_{t}^{\t}dt'+\int_{0}^{\t}dt'\int_{t'}^{\t}dt.
\eeq
In the first term we set $u=t$, $v=t'-t$, and in the second we set $u=t'$, $v=t-t'$ to obtain,
\bea
\Pi(y)&=&\frac{1}{m^{2}}\int d\h d\h'\brak{\phi_{0}}{\h'}\brak{\h'}{y}\brak{y}{\h}\brak{\h}{\phi_{0}}\int_{0}^{\t}du\int_{0}^{\t-u}dv\nonumber\\
&&\times\left\{\bra{\psi_{0}}\exp(i(H_{0}+\l \h')u)\hat p \delta(\hat x)\hat p\exp(-i(H_{0}+\l \h)(u+v))\ket{\psi_{0}}\bra{0}\exp(iH_{0}v)\ket{0}\right.\nonumber\\
&&\left.+\bra{\psi_{0}}\exp(i(H_{0}+\l \h')(u+v))\hat p\delta(\hat x)\hat p \exp(-i(H_{0}+\l \h)u)\ket{\psi_{0}} \bra{0}\exp(-iH_{0}v)\ket{0}\right\}\nonumber\\
\eea
Since we take the time $\t$ to be large, we can extend the upper limits of the integrals to infinity. The integral over $v$ can then be carried out, to give,
\bea
\Pi(y)&=&\int d\h d\h'\brak{\phi_{0}}{\h'}\brak{\h'}{y}\brak{y}{\h}\brak{h}{\phi_{0}}\nonumber\\
&&\times\frac{(-1)}{2m}\int_{0}^{\infty}du \bra{\psi_{u}}e^{i\l \h'u}\big(\hat p \delta(\hat x)+\delta(\hat x) \hat p\big) e^{-i\l \h u}\ket{\psi_{u}}\nonumber\\
\Pi(y)&=&\frac{(-1)}{2m}\int_{0}^{\infty}du |\Phi(y,u)|^{2}\bra{\psi_{u}}\big(\hat p \delta(\hat x)+\delta(\hat x) \hat p\big) \ket{\psi_{u}}\nonumber\\
&=&\int_{0}^{\infty} dt |\Phi(y,t)|^{2} J(t)
\label{3.8}
\eea
where
 \beq
\Phi(y,t)=\int d\h\brak{y}{\h}\brak{\h}{\phi_{0}}e^{-i\l \h t}=\bra{y}e^{-i\lambda H_{c}t}\ket{\phi_{0}}
\eeq
is the wavefunction of the clock, and $J(t)$ is the current, Eq.(\ref{cur}).

This form shows that, in the weak coupling limit, our arrival time probability
distribution yields the current, but smeared with a function depending on the clock state.
We thus get agreement with the expected result, Eq.(\ref{best}).
Note that the physical quantity measured, the current, is not
affected by the form of the clock Hamiltonian.

Although the form Eq.(\ref{3.8}) holds for a wide class of clock Hamiltonians, not all choices make for equally good clocks. To further restrict the coupling $H_{c}$ we require that different arrival times may be distinguished up to some accuracy $\delta t$. For this to be the case we require that the clock wavefunctions corresponding to different arrival times are approximately orthogonal, so that
\beq
\int dy \Phi^{*}(y,t') \Phi(y,t)\approx\begin{cases}1 &\mbox{ if } t\approx t'\\
0 &\mbox{ otherwise}\end{cases}
\eeq
We easily see that,
\bea
\int dy \Phi^{*}(y,t') \Phi(y,t)&=&\int d\h d\h'dy \brak{\phi_{0}}{\h'}\brak{\h'}{y}\brak{y}{\h}\brak{\h}{\phi_{0}}e^{i\l (t' \h'-t \h)}\nonumber\\
&=&\int d\h|\phi_{0}(\h)|^{2}e^{-i\l \h\delta t}\label{acc}
\eea
where $\delta t=t-t'$. Clearly this expression is equal to 1 if $\delta t=0$. Suppose now that $|\phi_{0}(\h)|^{2}$ is peaked around some value $\h_{0}$ with width $\sigma_{\h}$. This integral will approximately vanish if
\beq
\l  \sigma_{\h} \delta t>>1,
\eeq
and so the resolution of the clock is given by $1/\l\sigma_{\h}$.
The relationship between $t$ and the pointer variable $y$
will depend on the specific model.
It is easily seen that a clock with good characteristics may be
obtained using, for example, a free particle with a Gaussian
initial state. But clocks with more general Hamiltonians
can also be useful if they evolve an initial Gaussian along
an approximately classical path (as many Hamiltonians do).
See Refs.\cite{clock,clock2,Hartleclock,clockchapter}
for further discussion of clock characteristics.

For the special case $H_{c}=\hat p_{y}$, $\ket{\h}=\ket{p_{y}}$,  the expression for the arrival time distribution simplifies, since
\beq
\Phi(y,t)=\int \frac{dp_{y}}{\sqrt{2\pi}}e^{ip_{y}(y-\l t)}\tilde\phi_{0}(p_{y})=\phi_{0}(y-\l t)
\eeq
The time is related to $y$ by $ t = y/\lambda$ and the
expected form Eq.(\ref{best}) then becomes a simple
convolution.

\subsection{Strong Coupling Regime}
\subsubsection{Special Case: $H_{c}=\hat p_{y}$}

We now turn to the limit of strong coupling between the particle and clock. The analysis of the case of general clock Hamiltonian is rather subtle, so before we tackle this we first examine the special case where the clock Hamiltonian is linear in the momentum. That is, we have,
\beq
H_{c}=\hat p_{y}=\int dp_{y}\;p_{y}\ket{p_{y}}\bra{p_{y}}.
\eeq
We start from Eq.(\ref{atstart}), and insert a complete set of momentum states for the particle, $p$ to obtain,
\bea
\Psi(x,y,\t)&=&\frac{1}{m}\int \frac{dp_{y} dp}{\sqrt{2\pi}}\brak{y}{p_{y}}\tilde\phi_{0}(p_{y})\exp(-i(E+\l p_{y})\t)p\brak{p}{\psi_{0}}\nonumber\\
&&\times\int_{0}^{\t}dt \bra{x}\exp(-i(H_{0}-\l p_{y}\theta(-\hat x))t)\ket{0} \exp(iEt),\nonumber\\
\eea
where $\tilde \phi_{0}(p_{y})$ is the initial momentum space wavefunction of the clock, and $E=p^{2}/2m$ is the kinetic energy of the particle.
Note the appearance of the momentum $p$ in the integrand.
The expression involving the integral over $t$ has been computed previously using the final crossing PDX \cite{Hal2, HaYe1}. In the limit $\t\to\infty$ it is given by,
\bea
\int_{0}^{\infty}dt \bra{x}\exp(-i(H_{0}-\l p_{y}\theta(-\hat x))t)\ket{0} \exp(iEt)
&=&\sqrt{\frac{2m}{\l p_{y}}}\exp\left(-ix\sqrt{2m(E+\l p_{y})}\right)\nonumber.
\eea

We can now write our probability distribution for $y$. Carrying out the $x$ integral we obtain,
\bea
\Pi(y)&=&\int dp_{y} dp_{y}'dpdp'\tilde \phi_{0}^{*}(p_{y}')\tilde \phi_{0}(p_{y})\brak{p_{y}'}{y}\brak{y}{p_{y}}\exp(-i\l(p_{y}-p_{y}')\t)\nonumber\\
&&\times \frac{pp'}{m}\brak{\psi_{0}}{p'}\brak{p}{\psi_{0}}\exp(-i(E-E')\t)\nonumber\\
&&\times \frac{2}{\sqrt{\l^{2}p_{y} p_{y}'}}\delta\left(\sqrt{2m(E+\l p_{y})}-\sqrt{2m(E'+\l p_{y}')}\right).\nonumber\\
\eea
Using the formula $\delta(f(x))=\delta(x)/f'(0)$, we can carry out the $p_{y}'$ integral to give,
\bea
\Pi(y)
&\approx&\int dp_{y} dpdp'|\tilde \phi_{0}(p_{y})|^{2}\frac{pp'}{m^{2}}\brak{\psi_{0}}{p'}\brak{p}{\psi_{0}}\exp\left(-i\frac{(E-E')}{\l}y\right)\frac{2}{\l}\sqrt{\frac{2m}{\l p_{y}}}\label{3.17}\\
&=&\int dp_{y}|\tilde \phi_{0}(p_{y})|^{2}\frac{2}{ m^{2}}\sqrt{\frac{2m}{\l p_{y}}}\bra{\psi_{0}}\exp\left(iH_{0}\frac{y}{\l}\right)\hat p\delta(\hat x)\hat p\exp\left(-iH_{0}\frac{y}{\l}\right)\ket{\psi_{0}}\nonumber\\
&=& A\bra{\psi_{0}}\exp\left(iH_{0}\frac{y}{\l}\right)\hat p\delta(\hat x)\hat p\exp\left(-iH_{0}\frac{y}{\l}\right)\ket{\psi_{0}}
\eea
where $A$ is some constant whose explicit form is not required and we have used the fact that $E<<\l p_{y}$ to approximate,
\beq
\phi\left(p_{y}+\frac{E-E'}{\l}\right)\approx\phi(p_{y}).
\eeq
We see therefore that in this limit the probability of
finding the clock at a position $y$ is given by the kinetic
energy density of the system at the time $t=y/\l$, in agreement
with Eqs.(\ref{Zeno}) and (\ref{normalized}).

Note that there is no response function involved in this case, as
one might have expected from the general form Eq.(\ref{best}).
A similar feature was noted in the complex potential model of
Ref.\cite{Hal2}). It seems likely that this is because the strong
measurement prevents the particle from leaving $x>0$ until the
last moment, so that
the response function $R(t,s)$ is effectively a delta-function
concentrated around the latest time.

\subsubsection{General Case}

As well as the approximations valid for $E<<\l p_{y}$, the key to the analysis in the special case presented above is that the position space eigenfunction of the clock Hamiltonian with eigenvalue $p_{y}$ takes the simple form,
\beq
\brak{y}{p_{y}}=\frac{1}{\sqrt{2\pi}}\exp(i y p_{y}).
\eeq
This greatly simplifies the resulting calculation. For the case of a more general clock Hamiltonian, the eigenstates will not have this simple form. Instead we make a standard WKB approximation for the eigenstates of the clock,
\beq
\brak{y}{\h}=C(y,\h)\exp(i S(y,\h))
\eeq
where $S(y,\h)$ is the Hamilton Jacobi function of the clock at fixed energy.  This means Eq.(\ref{3.17}) becomes,
\bea
\Pi(y)
&\approx&\int d\h dpdp'|\brak{\h}{\phi_{0}}|^{2}\frac{pp'}{m^{2}}\brak{\psi_{0}}{p'}\brak{p}{\psi_{0}}\brak{\h+\frac{(E-E')}{\l}}{y}\brak{y}{\h}\frac{2}{\l}\sqrt{\frac{2m}{\l \h}}\nonumber\\
&\approx&\int d\h|\brak{\h}{\phi_{0}}|^{2}\frac{2}{ m^{2}}\sqrt{\frac{2m}{\l \h}}|C(y,\h)|^{2}\nonumber\\
&&\times\bra{\psi_{0}}\exp\left(iH_{0}\frac{1}{\l}\frac{\partial S(y,\h)}{\partial \e}\right)\hat p\delta(\hat x)\hat p\exp\left(-iH_{0}\frac{1}{\l}\frac{\partial S(y,\h)}{\partial \h} \right)\ket{\psi_{0}}\label{zfin}
\label{Pistrong}
\eea
where we have used,
\beq
\brak{\h+\frac{(E-E')}{\l}}{y}\brak{y}{\h}\approx |C(y,\h)|^{2}\exp\left(-i\frac{(E-E')}{\l} \frac{\partial S(y,\h)}{\partial \h}\right),
\eeq
which is valid for $E-E'<<\l \h$.

We now suppose that the clock state is a simple Gaussian in $y$,
or equivalently in $p_y$. It follows that it will be peaked in
$\h$ about some value $\h_0$. This means that the integral over $\h_0$ in
Eq.(\ref{Pistrong}) may be carried out. The result for $\Pi(t)$ will again be
proportional to the kinetic energy density, of the form
Eq.(\ref{Zeno}), where the relationship between $t$ and the
pointer variable $y$ is defined by the equation
\beq
t=\frac{1}{\l}\frac{\partial S(y,\h_0)}{\partial \h}
\eeq
as one might expect from Hamilton-Jacobi theory \cite{Goldstein}.
Hence the arrival time distribution has the expected general form,
Eq.(\ref{Zeno}) (and therefore Eq.(\ref{normalized}) also holds),
but the precise definition of the time variable depends on the
properties of the clock.

\section{Dwell Times}\label{sec4}

\begin{figure}[htbp] 
   \centering
   \includegraphics[width=5in]{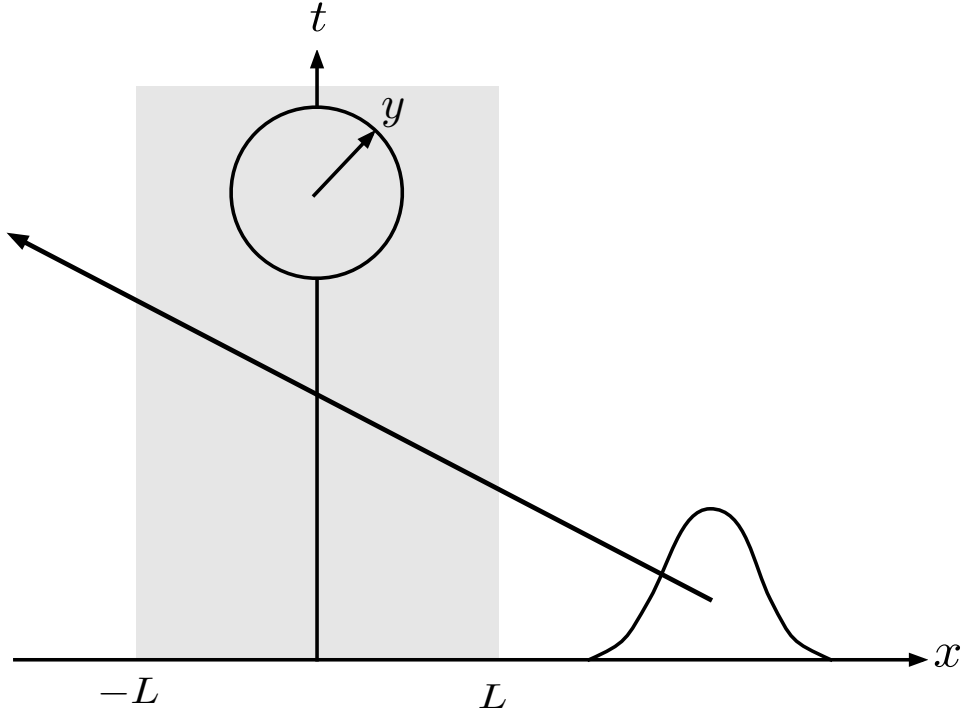}
   \caption{The dwell time problem, defined using a model clock. The clock runs while the particle is in the region $[-L,L]$.}
   \label{fig:3}
\end{figure}

We now turn to the related issue of dwell times. Here the aim is
to measure the time spent by the particle in a given region of
space which, for simplicity, we take to be the region $[-L,L]$.
This is portrayed in Figure (\ref{fig:3}). In
this section we will work exclusively in the weak coupling regime
where $E>>\l \h$.

\begin{figure}[htbp] 
   \centering
   \includegraphics[width=5in]{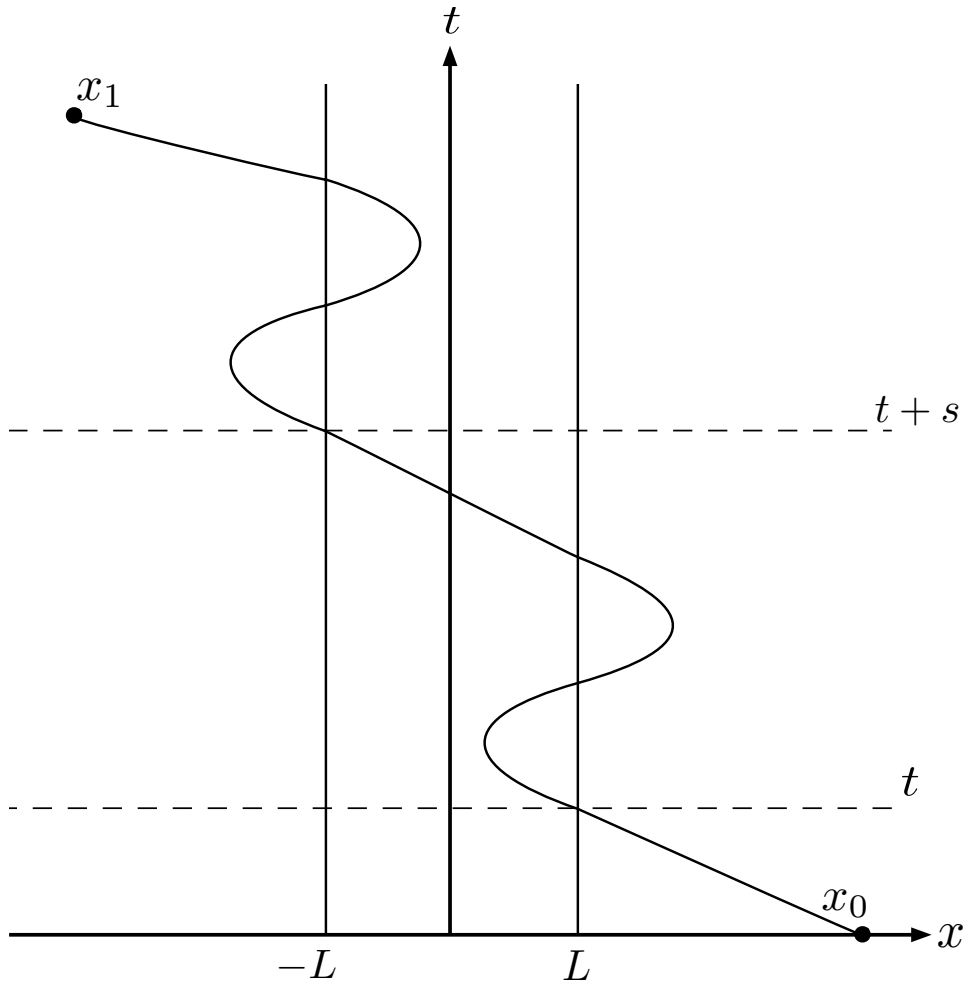}
   \caption{The PDX used for the dwell time problem: Paths from $x_{0}>0$ to $x_{1}<0$ have a first crossing of $x=L$ at a time $t$ and a first crossing of $x=-L$ at a time $t+s$. }
   \label{fig:4}
\end{figure}

The starting point is the final state of the particle plus clock,
Eq.(\ref{psifinal}), which we write as
\beq
\Psi (x,y, \tau )=\bra{x,y}e^{-i(H_{0}+\l H_{c}\chi(\hat x))\t}\ket{\Psi_{0}}.
\eeq
where $\chi(\hat x)=\theta(\hat x+L)\theta(L-\hat x)$.
We wish to re-express this using the path decomposition in a similar way to Eq.(\ref{atstart}).
For this case we need a PDX which is more general than the one
used for the arrival time, since there are now crossings of two
surfaces.
One way to proceed is to use the path integral expression for the first crossing of $x=L$ and
$x=-L$, which is
\bea
\Psi (x,y, \tau)&=&\frac{1}{m^{2}}\int d\h\brak{\h}{\phi_{0}}\brak{y}{\h}\int_{0}^{\infty}ds\int_{-\infty}^{\t-s}dt \bra{x}\exp(-i(H_{0}+\l\chi(\hat x)\h)(\t-s-t))\ket{-L}\nonumber\\
&&\bra{-L}\hat p\exp(-i(H_{0}+\l\chi(\hat x)\h)s)\ket{L}\bra{L}\hat p\exp(-iH_{0}t)\ket{\psi_{0}},\label{4.2}
\eea
This is shown in Figure (\ref{fig:4}).
However there are other choices we could make.
We could consider the first crossing of $x=L$ and the last crossing of $x=-L$ for example.
In the semiclassical limit these choices lead to equivalent expressions for the dwell time.
It would be interesting to explore what differences do arise in other regimes. This will be addressed elsewhere.

It will prove more useful to work with the wavefunction in position space for the clock and momentum space for the particle. Changing to this representation, and making use of the semiclassical approximation, Eq.(\ref{PDX2}), we obtain,
\bea
\brak{p,y}{\Psi_{\t}}&\approx&\frac{1}{m^{2}}\int d\h\brak{y}{\h}\brak{\h}{\phi_{0}}\int_{0}^{\infty}ds\int_{-\infty}^{\t-s}dt \bra{p}\exp(-iH_{0}(\t-s-t))\ket{-L}\nonumber\\
&\times&\bra{-L}\hat p\exp(-i(H_{0}+\l \h)s)\ket{L}\bra{L}\hat p\exp(-iH_{0}t)\ket{\psi_{0}}
\eea
This is a semiclassical version of the PDX for first crossing of $x=L$ and $x=-L$.
Now we make the standard scattering approximation of letting the upper limit of the integral over $t$ go to infinity. This means we can carry out the $t$ and $s$ integrals to obtain,
\bea
\brak{p,y}{\Psi_{\t}}&\approx&\int d\h\brak{y}{\h}\brak{\h}{\phi_{0}} \exp(-iE\t)\exp(-i2Lm\l \h/|p|)\brak{p}{\psi_{0}},
\eea
where we have used the standard integral \cite{HaYe1},
\beq
\int_{0}^{\infty} ds \bra{x}\exp(-iH_{0}s)\hat p\ket{0}e^{iEs}=m\exp(i|x|\sqrt{2mE})
\eeq
and
\beq
\int_{-\infty}^{\infty}\frac{dt}{2\pi}\bra{x}\hat p \exp(-i(H_{0}-E)t)\ket{\psi_{0}}=\bra{x}\hat p \delta(H_{0}-E)\ket{\psi_{0}}=m\bra{x}\delta(\hat p-p)\ket{\psi_{0}}
\eeq
Here, we have neglected the term involving $\delta(\hat p+p)$ since this corresponds to reflection, which will be negligible in this semiclassical limit. We have also use the fact that $E\gg\l \h$ to approximate,
\beq
\exp\left(i2L\sqrt{2m(E-\l\h)}\right)\approx\exp\left(i2L|p|-i2Lm\l\h/|p|\right).
\eeq
We therefore obtain the distribution for $y$ as,
\bea
\Pi(y)&=&\int dp \left|\brak{p,y}{\Psi_{\t}}\right|^{2}\nonumber\\
&=&\int d\h d\h'\brak{\phi_{0}}{\h'}\brak{\h'}{y}\brak{y}{\h}\brak{\h}{\phi_{0}}\int dp|\psi_{0}(p)|^{2}\exp\left(\frac{i2Lm\l}{|p|}(\h'-\h)\right)\nonumber\\
&=&\int dp|\psi_{0}(p)|^{2}|\Phi(y,2Lm/|p|)|^{2}
\eea
where,
\bea
\Phi(y,2Lm/|p|)&=&\int d\h\brak{y}{\h}\brak{\h}{\phi_{0}}\exp\left(\frac{-i2Lm\l}{|p|}\h\right)\nonumber\\
&=&\bra{y}\exp\left(\frac{-i2Lm\l}{|p|} H_{c}\right)\ket{\phi_{0}}
\eea
is the clock wavefunction.
We may rewrite this as
\beq
\Pi (y) = \int dt \ |\Phi(y,t)|^{2} \langle \psi_0 | \delta \left( t
-\frac { 2 m L } { | \hat p | } \right) | \psi_0 \rangle
\eeq
It is therefore of precisely the desired form,
Eqs.(\ref{dwellsemi}), (\ref{best}), with $ |\Phi(y,t)|^{2} $
playing the role of the response function. The discussion of clock
characteristics is then exactly the same as the arrival time case
discussed in Section 3.

\section{Conclusion}\label{con}

We have studied the arrival and dwell time problems defined using
a model clock with a reasonably general Hamiltonian. We found that
in the limit of weak particle-clock coupling, the time
of arrival probability distribution is given by the probability
current density Eq.(\ref{cur}), smeared with some function depending on the
initial clock wave function, Eq.(\ref{best}). This is expected semiclassically,
agrees with previous studies and is independent
the precise form of the clock Hamiltonian.

In the regime of strong coupling, we found that the arrival time
distribution is proportional to the kinetic energy density of the
particle, in agreement with earlier approaches using a complex
potential. The fact that two very different models give the same
result in this regime suggests that the form Eq.(\ref{normalized})
is the generic result in this regime, independent of the method
of measurement. It would be of interest to develop a general
argument to prove this. (See Ref.\cite{ech} for further discussion
of this regime).

For the case of dwell time, we have shown that the dwell time
distribution measured by our model clock may be written in terms
of the dwell time operator in semiclassical form, smeared
with some convolution function, Eqs.(\ref{dwellsemi}),
(\ref{best}).

In all of these cases, the precise form of the
clock Hamiltonian and clock initial state determine
the relationship between time $t$ and the pointer variable
$y$ and they determine the form of the response function
$R$ in the general form Eq.(\ref{best}). These are particularly
simple for the special case $H_C = \hat p_y $ explored
previously. However, what is
important is that, once the definition of the time variable is fixed,
the clock characteristics do
not effect the form of the underlying distributions -- the $\Pi
(s)$ in Eq.(\ref{best}). The $\Pi (s) $ are always one of the
general forms
Eqs.(\ref{cur}), (\ref{Zeno}) and (\ref{dwellsemi}), no matter what
the clock characteristics are.
This means that these general forms will always play
central
role, irrespective of how they are measured.

\bibliography{apssamp}

\end{document}